\documentclass[10pt]{article}
\usepackage[OE]{express}
\usepackage{wrapfig}
\usepackage{amsmath}
\usepackage{braket}

\begin{document}

\title{Causality and information transfer in simultaneously slow- and fast-light media}

\author{Jon D. Swaim and Ryan T. Glasser\authormark{*}}

\address{\authormark{1}Physics and Engineering Physics Department, Tulane University, New Orleans, LA 70118, USA}

\email{\authormark{*}rglasser@tulane.edu} 

\begin{abstract}
We demonstrate the simultaneous propagation of slow- and fast-light optical pulses in a four-wave mixing scheme using warm potassium vapor.  We show that when the system is tuned such that the input probe pulses exhibit slow-light group velocities and the generated pulses propagate with negative group velocities, the information velocity in the medium is nonetheless constrained to propagate at, or less than, $c$.  These results demonstrate that the transfer and copying of information on optical pulses to those with negative group velocities obeys information causality, in a manner that is reminiscent of a classical version of the no-cloning theorem.  Additionally, these results support the fundamental concept that points of non-analyticity on optical pulses correspond to carriers of new information.  
\end{abstract}


\bibliographystyle{osajnl}




\section{Introduction}

In 1905, Einstein unified several prevalent concepts in the theory of electromagnetism, elevating them into a single, far-reaching principle of relativity~\cite{Einstein1905SR, Poincare1902, Lorentz1898, Michelson1887, Maxwell1865, Galison2003}.  The special theory of relativity asserts that Lorentz invariance --  i.e., that the speed of light, $c$, is the same in every reference frame -- is a fundamental aspect of nature.  A consequence of this is that superluminal signaling is prevented in flat spacetime, which would otherwise violate the principle of causality and lead to enigmatic outcomes, such as the grandfather paradox.  The effects of relativity in general have been confirmed in a range of experiments, including the measurement of information velocity~\cite{Stenner2003, Stenner2005, Tomita2011}, the transverse Doppler effect~\cite{Ives1938}, time dilation of moving atomic clocks~\cite{Hafele1972} and, most recently, the direct detection of gravitational waves~\cite{Abbott2016}.  Additionally, information causality, or no-signaling, has since become a broad and important physical concept: one which coexists peacefully with the classical notion of locality, and at the same time has been shown to be an essential ingredient of nonlocal physical theories possessing intrinsic randomness and prohibiting the perfect cloning of arbitrary states (no-cloning), e.g., quantum mechanics~\cite{Masanes2006}.

\begin{figure}[b!]
\begin{center}
\includegraphics*[width=0.8\columnwidth]{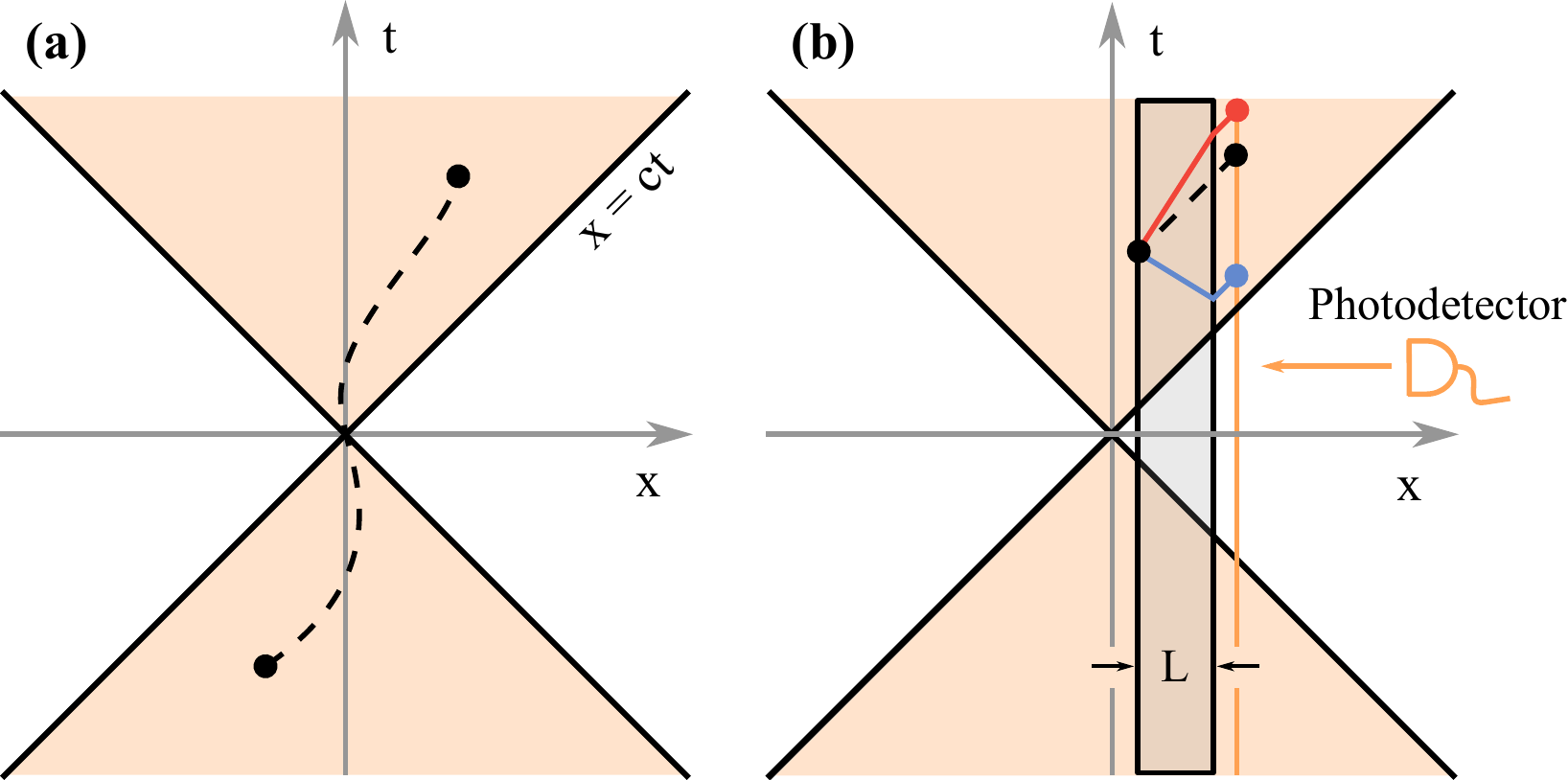}
\caption{\label{spacetime} Spacetime diagrams for information flow between two causally connected events through (a) free space and (b) a simultaneously fast- and slow-light medium of length $L$. In both cases, the black dotted line represents information flow.  The red and blue lines in (b) correspond to optical pulses with group velocities which are slow ($v_g < c$) and fast (negative, $v_g < 0$), respectively.}
\end{center}
\end{figure} 

\begin{figure*}[t!]
\begin{center}
\includegraphics*[width=\textwidth]{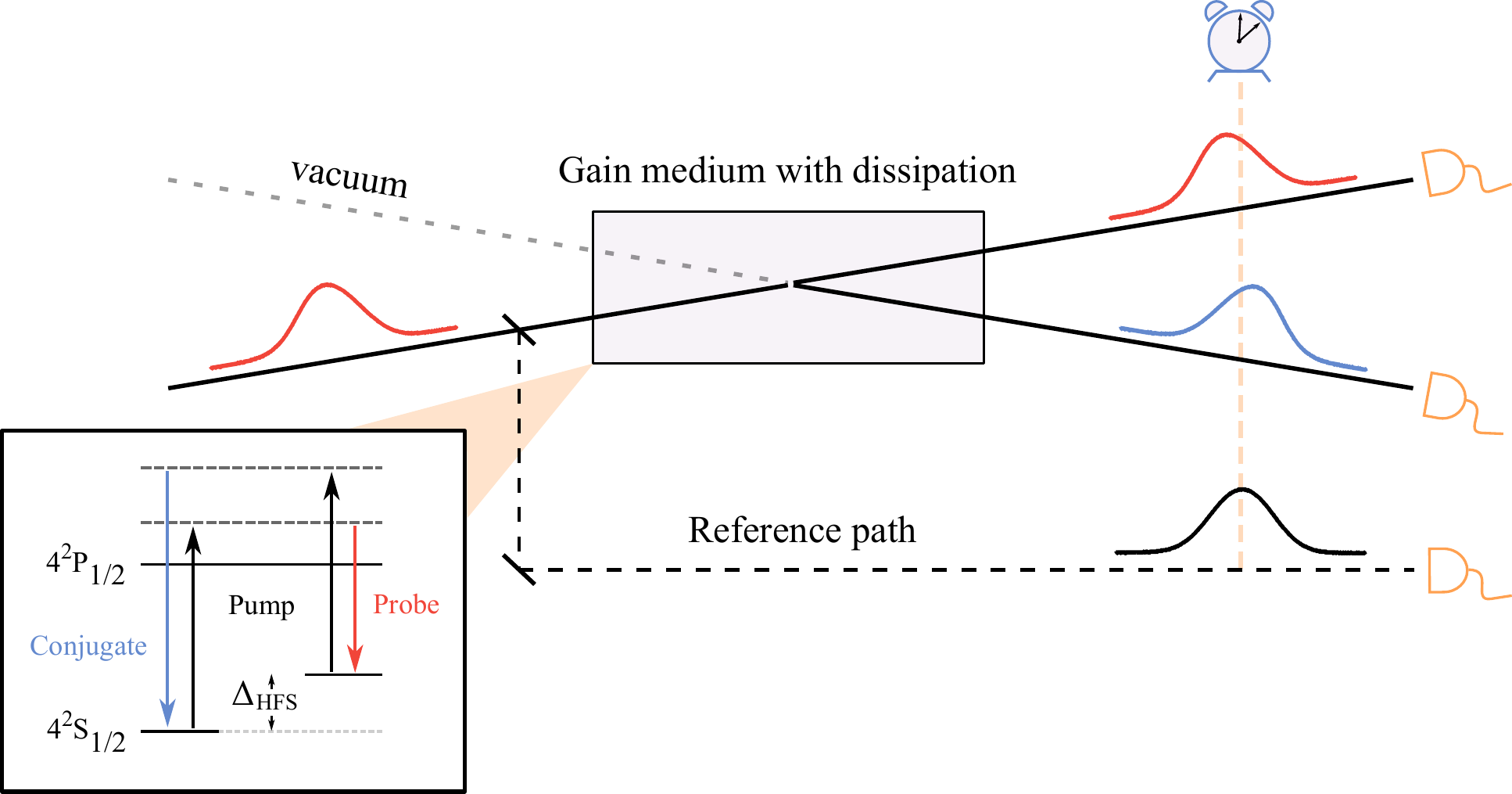}
\caption{\label{setup}Depiction of the four-wave mixing copier.  The probe input is seeded with an input pulse (shown in red), and the process generates a copy of the input (blue).  The arrival times of the pulses are compared with a reference of the speed of light in air (black).  The inset shows an energy level diagram which summarizes the four-wave mixing process.  The hyperfine ground state splitting in potassium is $\Delta_{\textrm{HFS}} = 462$ MHz.} 
\end{center}
\end{figure*}

The concept of superluminal pulse propagation, wherein optical pulses can propagate through dispersive media with group velocities greater than $c$, or even negative, has therefore stimulated much interest in causality over the years~\cite{Chu1982, Steinberg1994, Garrison1998, Wang2000, Kuzmich2001, DeCarvalho2002, Stenner2003, Nimtz2003, Solli2004, Stenner2005, Winful2006, Bigelow2006, Bianucci2008, Boyd2009, Tomita2011, Suzuki2013, Clark2014, Tomita2014, Macke2016, Dorrah2016, Amano2016, Asano2016}.  It is accepted that, while the envelop of a pulse can be modified such that the peak appears to travel faster than $c$, the $\emph{information}$ is a fundamental quantity bound to propagate at $c$ or slower~\cite{Brillouin1960, Garrison1998, Stenner2003, Parker2004, Stenner2005}.  In this view, the carriers of information are points of non-analyticity, and the dispersion-free nature of such off-resonant frequency components ensures a causal relationship upon transmission through the medium.  There is, therefore, a differentiation between the group velocity of a pulse $v_g$ and the information velocity $v_i$.  While the velocity of information has been investigated independently in the fast- and slow-light regimes, studies of information causality in a system exhibiting simultaneous slow- and fast-light propagation, as well as the transfer and copying of information to fast-light pulses, have not been realized.  Systems which are able to generate slow and fast light simultaneously offer great practical advantages, both in the pure and applied sciences, with the latter leading to implementations of all-optical information processing~\cite{Bianucci2008, Macke2016} and ultra-sensitive interferometry~\cite{Boyd2009}.   

In this paper, we examine information causality within the previously unexplored framework of copying and transferring information.  We report on a system which can be simultaneously slow and fast over a single mode, and demonstrate a four-wave mixing (4WM) copier which generates a copy of information (in the form of a step discontinuity) in a distinct optical mode, and deletes the original via absorption, resulting in information transfer between the two modes.  In agreement with information causality, we find that the information velocity is always less than $c$ when copying and transferring information, independent of the group velocity time shifts described above.  Moreover, we find that superluminal propagation can destroy information content, in a manner which is reminiscent of a classical version of the no-cloning theorem.  Our results are in agreement with those recently demonstrated involving the propagation of quantum mutual information through fast-light media~\cite{Clark2014}.  

In Fig.~\ref{spacetime}, we illustrate the general concept of causality and its relationship to transmitting information via light signals using spacetime diagrams.  In the free-space scenario, information flows between two events at a speed less than or equal to $c$, where the two events are connected through a worldline [dashed line in Fig.~\ref{spacetime}(a)].  The passage of information within spacetime regions limited by $x^2 \leq c^2 t^2$ ensures that the two events are causally connected.  

Let us now consider information passing through the medium in Fig.~\ref{spacetime}(b), which takes as its input an optical pulse traveling at $c$, and then generates two optical pulses, one of which is slow ($v_g < c$, red curve), and the other which is fast  (with a negative group velocity, $v_g < 0$, blue curve).  If we place a detector after the medium (orange line), it would seem given this simplified thought experiment that the fast pulse can arrive at the detector at a time before it was created.  However, since different frequency components of the pulse experience different dispersion, the medium acts as a filter and reshapes the pulse.  As a consequence, the incident and transmitted peaks, for example, are no longer causally connected~\cite{DeCarvalho2002}.  This realization motivated theoretical and experimental investigations of the idea that, in general, the information velocity of a signal is distinct from its group velocity.

To investigate this phenomenon in the context of information copying and transfer, we utilize a 4WM-based information copier with built-in loss and dispersion.  The copier is based on the double-$\Lambda$ scheme described previously in Refs.~\cite{McCormick2007, Zlatkovic2016}, and is laid out schematically in Fig.~\ref{setup}.  In this scheme, a gas of alkali atoms (in the current experiment, $^{39}K$) is strongly pumped near the D1 line, and a detuned probe pulse (shown in red) is injected into the medium.  The process amplifies the injected probe, and generates a second, conjugate pulse (blue) within the medium, with the probe and conjugate symmetrically detuned from the pump by approximately $\Delta_{\textrm{HFS}}$, as shown in the inset of Fig.~\ref{setup}.  As a result, any information imprinted on the original input pulse will be copied  into the conjugate.  Moreover, due to the relatively small ground state splitting and significant Doppler broadening in potassium, the configuration can be arranged such that the input probe pulse is strongly absorbed.  In this way, the setup can (i) generate the copy and (ii) erase the original via absorption, resulting in information transfer to an optical mode which is distinct in both frequency and space.

As discussed above, the transmitted pulses can be delayed or advanced as a result of dispersion, and under certain circumstances the time shifts can be achieved with minimal distortion~\cite{Macke2003}.  Whether the pulses are superluminal or subluminal depends on the sign of the medium's group index, which, experimentally, is controlled by varying the frequency of the probe with respect to the four-wave mixing resonance (i.e., the two-photon detuning $\delta$)~\cite{Glasser2012}. Since the probe and conjugate resonances are shifted in frequency with respect to each other, there exists an intermediate frequency (and therefore a pulse width $\tau$) where one of the outputs is superluminal while the other is subluminal.  However, in this regime, the time shifts are relatively small~\cite{Glasser2012}, as both outputs are tuned close to the transition between superluminality and subluminality.  We have found that greater control can be achieved by utilizing spatial dispersion, which is intrinsic in our system due to the alignment of the optical beams required by phase matching \cite{Glasser12}.  Spatial dispersion results in a spatially-varying group index, allowing different group velocities of interest to be selected by filtering the mode with an aperture.  Furthermore, when the frequency and pulse width are chosen so that the time shift associated with temporal dispersion is minimized, the optical pulse can exhibit both fast and slow light, simultaneously, over a single mode.

\section{Methods}

In the experiments, approximately 400 mW of coherent, continuous-wave light from a Ti:Sapph laser is collimated to an elliptical spot ($\sim 0.8$ mm $\times$ $0.6$ mm) and passed through an 80 mm, anti-reflection-coated potassium vapor cell.  This strong pump beam is detuned to the blue side of the Dopper-broadened D1 absorption profile by $\Delta$.  The input probe pulse is generated by double-passing a portion of the light through an acousto-optic modulator, which is driven at a variable frequency and modulated using an arbitrary wave generator to achieve a desired pulse shape.  The pump and probe ($1/e^2$ diameter of $\sim 670$ $\mu$m) beams are combined on a polarizing beam splitter so as to overlap in the center of the cell at a small angle ($\sim 0.1^\circ$).  The vapor cell is heated to a temperature of approximately $110$ $^\circ$C, and the average pulse power of the input probe is set to be approximately $10\, \mu$W.  After filtering out the pump light using a Glan-Taylor polarizer, the probe and conjugate pulses are spatially filtered using irises, directed to amplified photodetectors, and then analyzed on an oscilloscope.  For each experiment, the reference pulse is obtained by redirecting the light around the cell using a flip-mirror, which results in a negligible change in path length (i.e., the time shift is within the statistical error). 

\section{Results}

\subsection{Simultaneous generation of slow and fast light}

\begin{figure}[t!]
\begin{center}
\includegraphics*[width=0.8\columnwidth]{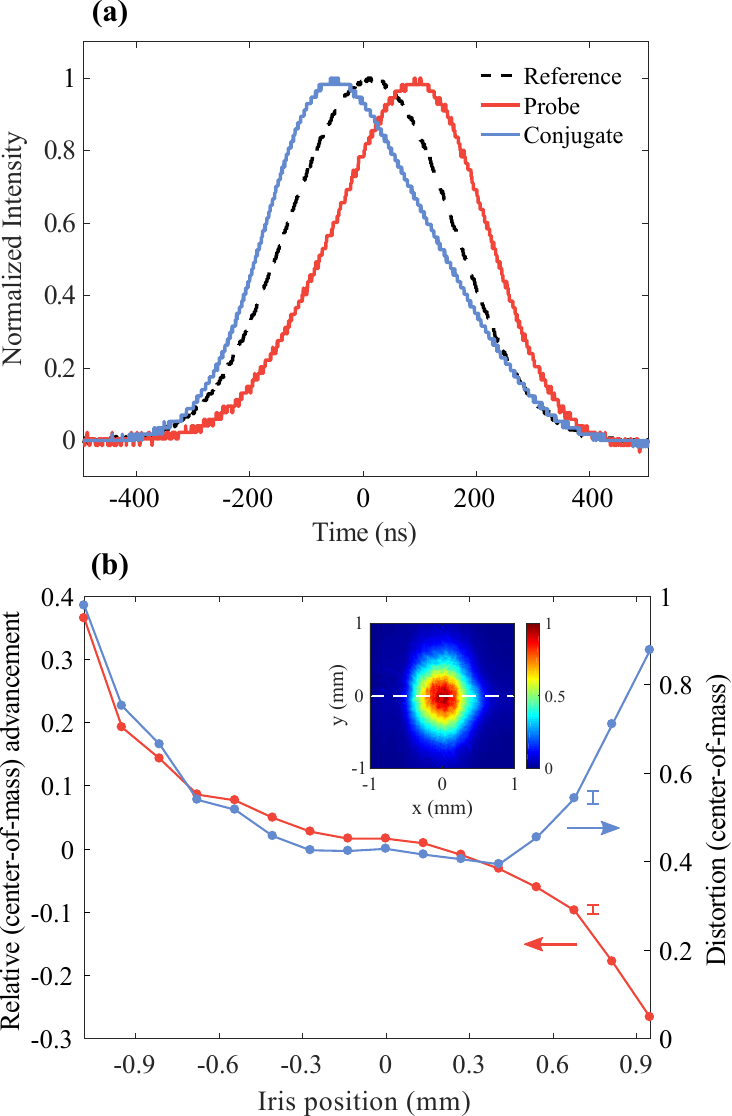}
\caption{\label{fastslow}Simultaneous slow and fast light.  (a)  Normalized intensities of detected pulses when the medium is tuned so that the probe and conjugate pulses are delayed and advanced, respectively, with respect to the reference pulse.  (b)  Relative (center-of-mass) advancement and calculated distortion as a function of spatial position over the conjugate mode.  Pulses are advanced for $x < 0$, delayed for $x > 0$ and  travel approximately at $c$ near $x = 0$.  The inset shows a normalized intensity image of the conjugate mode, with the dotted line indicating the position of the iris used in the measurement.}
\end{center}
\end{figure}

\begin{figure*}[t!]
\begin{center}
\includegraphics*[width=\textwidth]{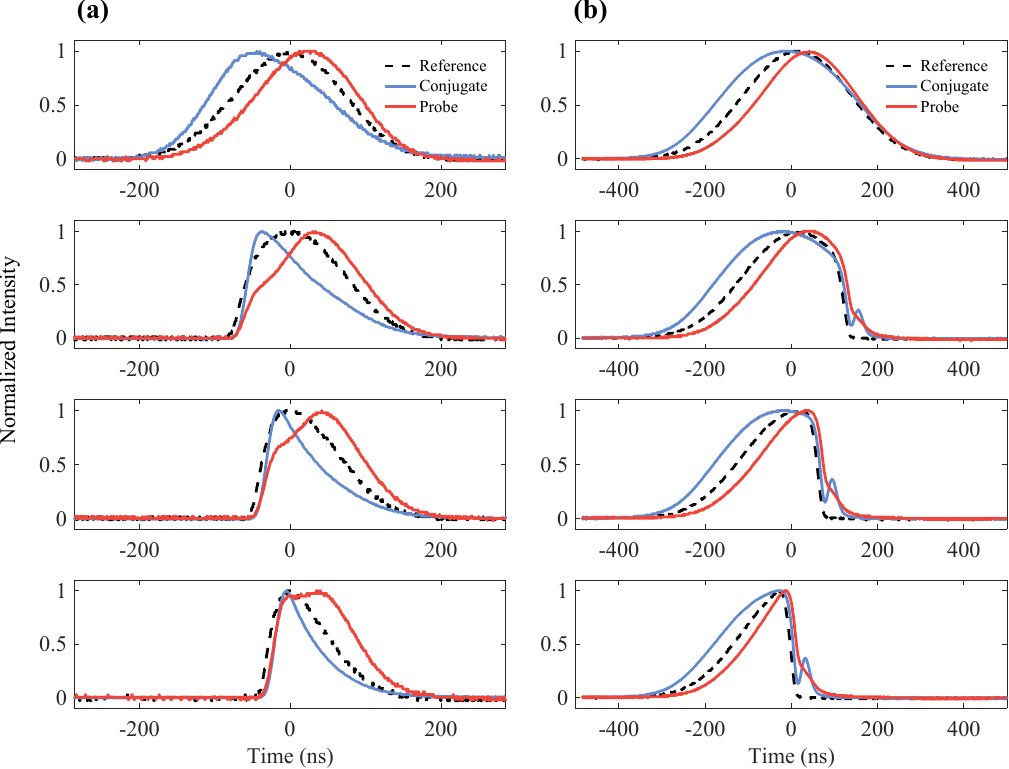}
\caption{\label{fastslowinfo}Arrival of new information in simultaneously slow- and fast-light media.  (a) Normalized pulse intensities when the new information is placed on the leading edge of the pulse.    The probe and conjugate modes are delayed and advanced by 14\% and 5\%, respectively.  (b)  Normalized pulse intensities when the new information is placed on the trailing edge of the pulse.  The delay and advancement are 11\% and 8\%, respectively.  In both (a) and (b), the discontinuity is displaced from the peak of the input pulse by $\tau/2$, $\tau/4$ or coincident with the peak (bottom row). Example pulses without the (approximately) non-analytic point are shown in the top row.}
\end{center}
\end{figure*}

In Fig.~\ref{fastslow} we show the results for simultaneous slow  and fast light in our system. The black dashed pulse in Fig.~\ref{fastslow}(a) is a reference propagating at the speed of light in air, which we obtain by redirecting the input pulse around the medium via a flip-mirror.  We optimize the system for simultaneously generating slow  and fast light by tuning the frequencies and filtering the spatial modes as described above, such that the overall effective gain of the four-wave mixer (sum of the outputs divided by the input) is $1.5\, \pm \,0.2$.  The experimental parameters (cell temperature, one- and two-photon detunings, etc.) for this measurement and all subsequent measurements are given in Table~\ref{parameters} at the end of the manuscript.  In this arrangement, the peaks of the probe and conjugate pulses, shown in red and blue, are shifted relative to the reference by $90$ ns (35\% of the original pulse width $\tau = 260$ ns) and $-60$ ns (23\%), respectively.  Considering the time shifts in terms of the center-of-mass (COM) of the pulse, rather than its peak, we find that the outputs are shifted by $22\%$ and $9\%$.  For the remainder of the manuscript, we will quantify the time shifts based on COM.  With the dispersive medium length of $L = 80$ mm, we calculate the group velocities $v_g = L/(\Delta t - L/c)$ for the probe and conjugate in this case to be $v_g = 5 \times 10^{-3}\, c$ and $v_g = - 11 \times 10^{-3}\, c$, respectively.  We reiterate that this result is for the case of two separate modes which have different frequencies and propagation directions, but the same polarizations.

Now we show that a similar result can be achieved over a single beam.  In the measurement presented in Fig.~\ref{fastslow}(b), we scan a small iris $\sim$ 1 mm in diameter horizontally over the conjugate mode (an image of the mode is shown in the inset), and compare the spatially filtered pulses' arrival times with that of the reference pulse.  We observe that a strong gradient emerges as a result of spatial dispersion, asymmetrically shifting the pulses by over $60 \%$ of the input pulse width, corresponding to nearly $100 \%$ based on the peak shifts.  Of particular interest is the fact that the profile is centered about a nearly zero time shift: i.e., superluminal for $x < 0$, subluminal for $x > 0$ and luminal near $x=0$.  We also include a calculation of the pulse disortion (see Ref.~\cite{Macke2003}), showing that the distortion is minimized for small time shifts and that a trade-off exists between dispersion and distortion at the far edges of the mode.  Due to this excess distortion (and nonlinear spatial dependence), in our experiments we optimize the generation of simultaneous slow  and fast light over the linear regime.  

While previous demonstrations of simultaneous slow  and fast light relied on either different polarizations~\cite{Bianucci2008, Macke2016} or different frequencies~\cite{Glasser2012}, in our case spatial dispersion generates both in a single mode.  In principle, this strategy could allow for the physics of nonlinear and quantum optics to be probed in the fast- and slow-light regimes at the same time, within a single ``shot." Another advantage of our setup is that both the probe and conjugate modes have profiles like that shown in Fig.~\ref{fastslow}(b), which allows for the fast- and slow-light outputs to be easily interchanged.  This is akin to alternating between absorption and gain, or flipping the sign of the dispersion that the input pulse experiences.  Lastly, we note that in the case of simultaneous slow and fast light, our advancements are comparable to those achieved previously~\cite{Bianucci2008}.

\subsection{Information velocity and the copying of optical information}

Now we turn to examining how the information content in the optical pulses is affected under conditions of fast and slow light, both from relativistic considerations and effects due to reshaping of the pulses.  In the earliest investigations, pulse fronts were considered as the true carriers of the information in optical pulses~\cite{Brillouin1960}. This suffers from the technical drawback that a pulse front is, by its very nature, much weaker than the rest of the pulse, and therefore non-ideal from an operational and measurement point of view.  Another approach is to impart a sharp discontinuity somewhere near the peak of the pulse, and take such a point of non-analyticity as the carrier of new information.  In practice, these features are analytic with a finite time constant $\tau_{\textrm{sig}}$, though they can approximate an ideal step function in the limit that $\tau_{\textrm{sig}} \ll \tau$.

Following the second approach, we measure the velocity of information in our simultaneously slow- and fast-light system, for a range of pulse widths, detunings and temporal positions of the discontinuity relative to the pulse peak.  Exemplary results are shown in Fig.~\ref{fastslowinfo}.  In the measurement of Fig.~\ref{fastslowinfo}(a), we tune our system to be fast and slow by $5\%$ and $14\%$, respectively, and introduce new information via a discontinuity on the leading edge of the input pulse.  We then compare the arrival times of the discontinuity on the probe and conjugate pulses with that of the reference.   In accordance with previous experiments on information velocity, we find that, despite the fact that the smooth peaks of the probe and conjugate pulses can propagate superluminally or subluminally, the new information is limited to propagate at $c$ or less.  This suggests that the group velocity is not always a meaningful concept with regards to the propagation of information.  These results also show that information causality holds when copying information between slow- and fast-light pulses, independent of the temporal position of the discontinuity.

Similar results can be achieved when the new information is placed on the trailing edge of the pulse.  In the result shown in Fig.~\ref{fastslowinfo}(b), we tune the 4WM copier to again produce simultaneous slow and fast light, using a different set of parameters, such that the pulses are advanced and delayed by $8\%$ and $11\%$.  In contrast to the leading edge case, here, we see qualitatively different behavior in the pulse reshaping, where the introduction of new information results in an additional peak following the discontinuity in the newly generated, copied pulse.  Despite these differences, the new information is limited to propagate at $c$ or less in both cases.

\subsection{Information transfer}

\begin{figure}[t!]
\begin{center}
\includegraphics*[width=0.8\columnwidth]{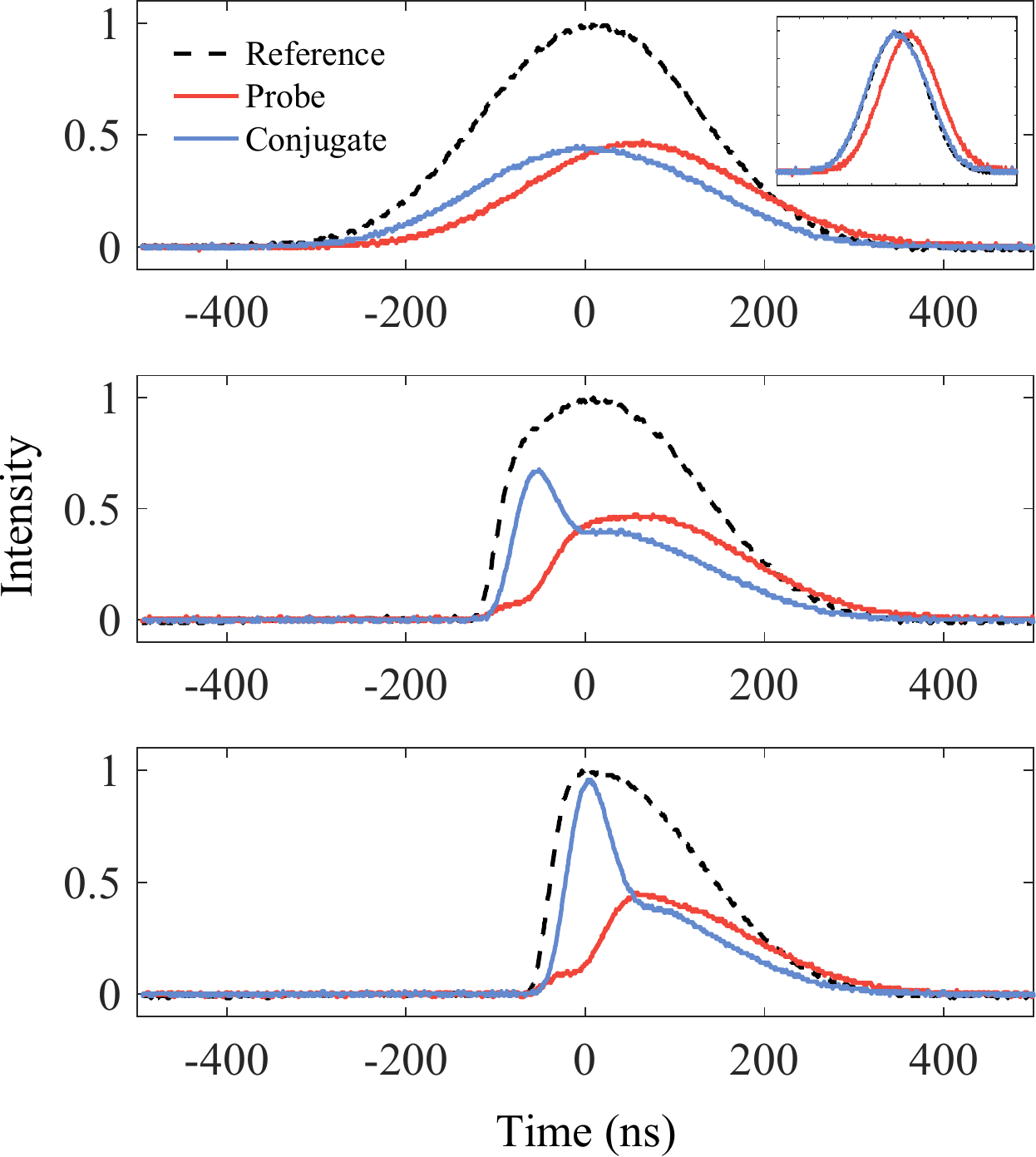}
\caption{\label{unitytransfer}Information transfer in the unity gain regime.  (Top) Pulse intensities when the probe is strongly absorbed and the effective total gain is $0.94 \pm 0.13$.  The inset displays the pulses normalized to the reference peak intensity, revealing that the conjugate channel is virtually dispersion-free and and probe is delayed by 26\%.  New information is placed on the leading edge of the input pulse, either $\tau/2$ away from the reference peak (middle) or at the peak (bottom).  The information signal is barely visible on the probe, whereas the relative signal intensity on the conjugate is nearly one, thus demonstrating information transfer to the conjugate.}
\end{center}
\end{figure}

\begin{figure*}[t!]
\begin{center}
\includegraphics*[width=\textwidth]{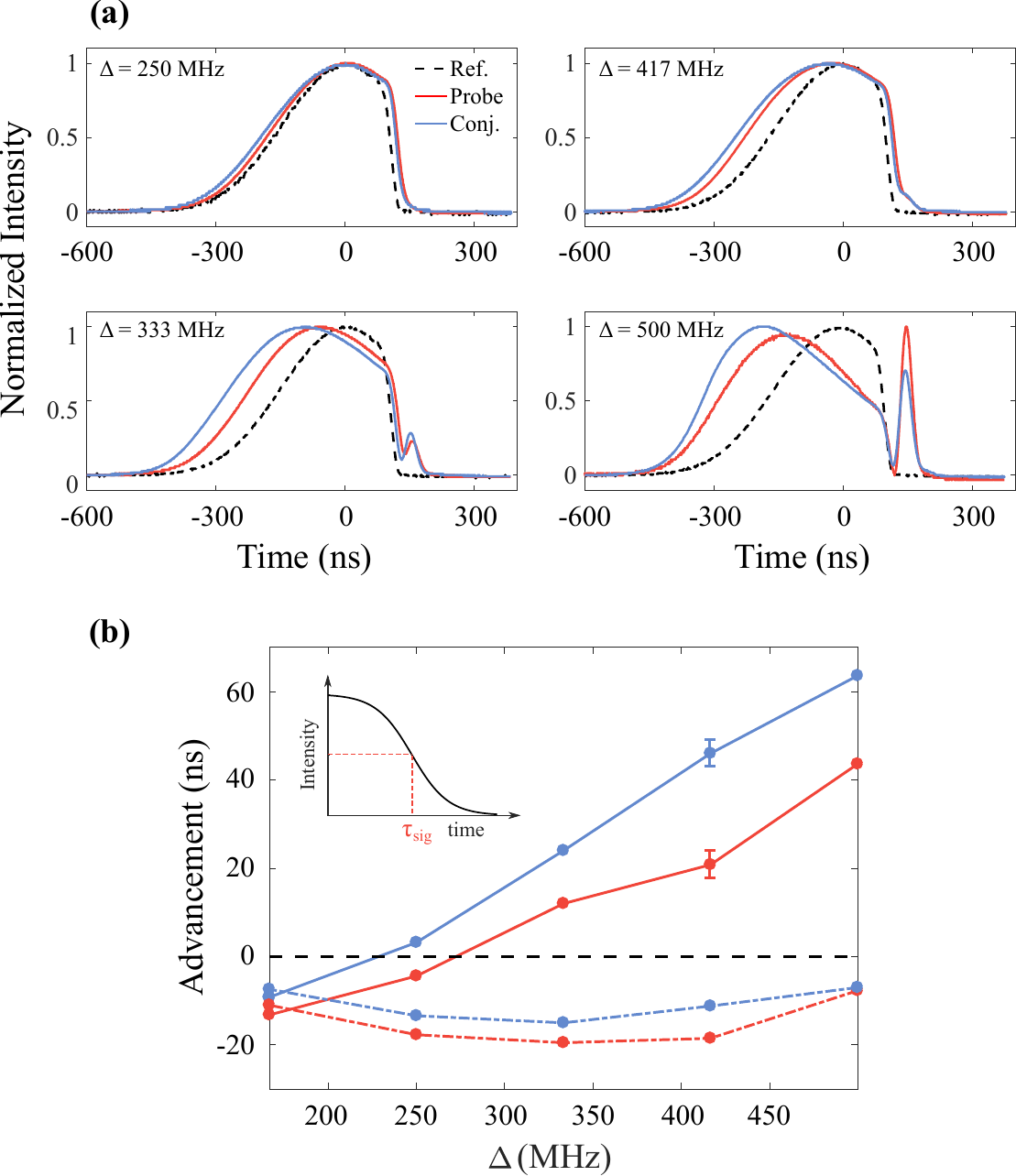}
\caption{\label{fastinfo} Information velocity and fast light.  (a) Normalized pulse intensities for various pump detunings, with the discontinuity placed on the trailing edges of the pulses.  Due to pulse reshaping, fast light is associated with a reduction in the intensity of the information signal, for both the probe and conjugate pulses. (b) Center-of-mass advancements and arrival times of the information signal, for the probe and conjugate data shown in (a).  The COM shifts are shown in solid lines for the probe (red) and conjugate (blue), and the dashed-dotted lines correspond to the relative arrival times of $\tau_{\textrm{sig}}$, the information signal.  Despite the group velocity advancements, both information signals always travels slower than the reference.}
\end{center}
\end{figure*}

The extension of these results to the transfer of information entails one final step: deletion of the information in the original (probe) pulse.  This step is achieved by tuning the frequencies of the beams such that the probe is detuned from the center of the Doppler-broadened resonances by $\approx$\,200 MHz, and thus experiences strong absorption.  We adjust the parameters so that the system operates with unity gain (with an effective, total gain of $0.94\, \pm\, 0.13$), and encounter a configuration where the dispersion experienced by the conjugate is inconsequential, and the probe is delayed by $26 \%$, with little distortion [top of Fig.~\ref{unitytransfer}].

Using these effects, we demonstrate irreversible, classical information transfer between the two optical modes.  From Fig.~\ref{unitytransfer}, we see that the intensity of the transferred information signal is approximately equal ($96\%$) to that in the reference pulse, and almost completely destroyed (only 9$\%$) in the absorbed probe pulse.  With a full optimization of parameters (most notably, the frequencies and bandwidths), it is possible that the probe signal strength could be reduced further.  Also, the tail end of the transmitted conjugate pulse is reshaped in this case, as the introduction of the discontinuity modifies the bandwidth, and hence the dispersion, which the pulse experiences.  

In the unity gain regime, our system behaves somewhat like a beamsplitter, with the addition that the frequency shift between the probe and conjugate resonances results in a delay between the two beamsplitter outputs.  A slow-light beamsplitter was realized previously using electromagnetically induced transparency in rubidium vapor~\cite{Xiao2008}.  In this case, the competition between gain and absorption inherent in the potassium-based 4WM scheme generates the dispersion, even with unity gain.

\subsection{Fast light and information}

To investigate the role of superluminality in our information copier, we adjust the setup so that both outputs are advanced, and then tune the advancement by varying the pump detuning ($\Delta$), keeping $\delta$ fixed.  In Fig.~\ref{fastinfo}(a), we show the resultant pulses for these measurements, which are taken in the amplifying regime, except for the last trace where the probe alone is slightly deamplified due to absorption, as discussed above.  Strong fast-light effects are evident for larger pump detunings, when the probe is positioned near the center of the Doppler-broadened absorption profile.  We show in Fig.~\ref{fastinfo}(b) the calculated COM time shifts for these pulses, as well as the relative arrival times of the information signal, $\Delta\tau_{\textrm{sig}}$, which are extracted based on the intensity time constants of these signals [see the inset of Fig.~\ref{fastinfo}(b)].  Despite the fact that the superluminal group velocities result in large advancements of the probe and conjugate pulses (up to 60 ns), the information signal always travels slower than $c$, with a delay which is on the order of 10 ns.  Furthermore, it clear from the pulses in Fig.~\ref{fastinfo}(a) that significant pulse reshaping accompanies the advancement, which has the effect of reducing the transmitted intensity of the information signal and destroying the information.  In other words, we observe a trade-off between superluminality and signal strength for both outputs.  In particular, when copying information to and from fast-light pulses, the trade-off occurs in an analogous way to a recent experiment involving quantum mutual information and fast light~\cite{Clark2014}, which is reminiscent of a classical version of the no-cloning theorem. 

\begin{table}[h!]
\centering
\caption{\label{parameters} Experimental parameters}
\bigskip
\begin{tabular}{cccccc}
Figure & T ($^\circ C$) & $\Delta$ (MHz) & $\delta$ (MHz) & $\tau$ (ns) & $\#$ of averages \\
\hline
3a & 109 & 666 & -8.5  & 260 & 128 \\
3b & 109 & 500 & -13.5 & 220 & 128 \\
4a & 108 & 250 & -54   & 130 & 16 \\
4b & 108 & 250 & -41.5 & 220 & 16 \\
5  & 108 & 666 & -2.3  & 220 & 16  \\
6  & 108 & *   & -26   & $\dagger$ &  128\\
\hline
\multicolumn{6}{l}{$*$ indicates this parameter varies in the measurement.} \\
\multicolumn{6}{l}{$\dagger$ $\tau = 260$ ns in the original Gaussian pulse.} \\
\end{tabular}
\label{parameters}
\end{table}

\section{Conclusions}

In summary, we have shown that the transportation of classical information between two optical modes is bounded by $c$, and that neither the pulse peak nor the COM reasonably describe the rate of information transfer resulting from the process.  Rather, the experiments described here reinforce the view that information is contained in points of non-analyticity, demonstrating that an all-optical information copier preserves information causality, regardless of sub- or superluminal group velocities.  In part, these results have been enabled by the fact that the system is highly tunable, in that both outputs can be simultaneously slow and fast over a single mode, with a maximum range of tuning corresponding to an entire pulse width.  Ultimately, the experiment is based on a straightforward optical setup, without cavities or cold atom ensembles, and the combined effects of the coupling between atomic transitions and absorption give rise to strong dispersion, even in the regime of unity gain, enabling the investigations elaborated upon here.

\bigskip

\section*{Funding}
We would like to thank the Louisiana State Board of Regents (Grant 073A-15) and Northrop Grumman $\emph{NG - NEXT}$ for generous funding which supported this work.

\bigskip



\end{document}